\documentclass[12pt,english]{article}
\usepackage[latin1]{inputenc}
\usepackage{color}
\makeatletter

\newcommand{\be}{\begin{equation}}
\newcommand{\ee}{\end{equation}}
\newcommand{\bea}{\begin{eqnarray}}
\newcommand{\eea}{\end{eqnarray}}
\newcommand{\ba}{\begin{array}}
\newcommand{\ea}{\end{array}}

\def\bbox{{\,\lower0.9pt\vbox{\hrule \hbox{\vrule height 0.2 cm
\hskip 0.2 cm \vrule height 0.2 cm}\hrule}\,}}
\newcommand{\dsl}{\pa \kern-0.5em /}

\newcommand{\Pslash}{{\bP} \hskip -.24truecm  / }

\def\bfn{\mbox{\boldmath $\nabla$}}

\usepackage{latexsym}
\usepackage{amsmath,,calrsfs}
\usepackage{amsfonts}
\usepackage{amssymb}
\usepackage{amscd}
\usepackage{bbm}
\usepackage{fancybox}
\usepackage{cite}
\usepackage{amsmath,amsfonts,amsbsy}
\usepackage{pstricks,pst-node}
\usepackage[small,bf,hang]{caption2}
\usepackage{graphicx}
\usepackage{epsfig}
\usepackage{psfrag}
\usepackage{comment}
\font\mybb=msbm10 at 12pt
\def\bb#1{\hbox{\mybb#1}}
\def\bZ {\bb{Z}}

\def\bN {\bb{N}}

\def\bC {\bb{C}}

\def\bX {\bb{X}}
\def\bP {\bb{P}}

\def\bU {\bb{U}}
\def\bV {\bb{V}}

\usepackage{babel}
\makeatother
\begin{document}
%%%%%%%%%%%%%%%% title page %%%%%%%%%%%%%%%%%%%%%%%%%%%%%%%%%%%%

%%%%%%%%%%%%%%%% title page %%%%%%%%%%%%%%%%%%%%%%%%%%%%%%%%%%
\begin{titlepage}

\begin{flushright}  UMTG-18 \\  DAMTP-2010-59
%$\hspace{2.1cm}{}$
\end{flushright}

\vfill

%\centerline{\Large{ DRAFT}}
%\vfill

\begin{center}
\baselineskip=16pt {\Large\bf Semionic Supersymmetric Solitons}

\vskip 2cm

{\large\bf Luca Mezincescu$^{\dagger, 1}$ and   Paul K.
Townsend$^{\star,2}$} \vskip 1cm {\small $^\dagger$ Department of Physics, University of Miami,\\
Coral Gables, FL 33124, USA 
\\}
\vspace{12pt}

{\small
$^\star$
Department of Applied Mathematics and Theoretical Physics\\
Centre for Mathematical Sciences, University of Cambridge\\
Wilberforce Road, Cambridge, CB3 0WA, UK\\
}

\end{center}
\vfill

\par
\begin{center}
{\bf ABSTRACT}
\end{center}
\begin{quote}

The Bogomolnyi vortex of the  ${\cal N}=2$ supersymmetric abelian-Higgs model in 2+1 dimensions  is shown to be a  ``semion'' of spin $1/4$.  Specifically, the  effective superparticle  action for one vortex  is shown to describe, upon quantization,  a parity self-dual centrally-charged 
`short'  supermultiplet  of  ``relativistic helicities''  $(-\tfrac{1}{4},-\tfrac{1}{4}, \tfrac{1}{4},\tfrac{1}{4})$.

\vfill \vfill \vfill \vfill \vfill \hrule width 5.cm \vskip 2.mm {\small \noindent $^1$
mezincescu@physics.miami.edu \\ \noindent $^2$ p.k.townsend@damtp.cam.ac.uk \\
}
\end{quote}
\end{titlepage}
%%%%%%%%%%%%%%%%%%%%%%%%%%%%%%%%%%%%%%

%%%%%%%%%%%%%%%%%%%%%%%%%%%%%%%%%%%%

\section{Introduction}
\setcounter{equation}{0}

There are a number of field-theory models in three spacetime dimensions (3D) in which non-perturbative excitations carry fractional spin and statistics, e.g. 
\cite{Wilczek:1983cy}. We refer the reader to the book by  Wilczek \cite{Wilczek:1990ik} for an introduction to the topic and a collection of early articles. The statistical phase angle $\theta$ and  the spin $s$ are related by $\theta  = 2\pi s$.   Fermions have spin $\tfrac{1}{2}+n$ for  integer $n$,  so ``half-fermions'' have spin $\tfrac{1}{4} + \tfrac{n}{2}$ and hence statistical 
phase $(\tfrac{1}{2} + n) \pi$.  The statistical phase of a bound state of two anyons of statistical phase $\theta$ is $4\theta$,  so the statistical phase of a bound state of two half-fermions is  a multiple of $2\pi$, which is equivalent to zero; in other words, the bound state is a boson. As bosons may condense, this makes half-fermions of potential relevance to high $T_c$ superconductivity.   In the context of relativistic field theory, half-fermions were initially  called ``quartions'' \cite{Volkov:1989qa} but are now generally called  ``semions'' \cite{Sorokin:1992sy,Sorokin:2002st}. In this context they have various special properties; one is that  two semions of spin $\tfrac{1}{4}$ and spin $\tfrac{3}{4}$ have a spin difference of $\tfrac{1}{2}$ and hence may appear together as the states of a single 
${\cal N}=1$ 3D  supermultiplet \cite{Sorokin:1992sy}; other aspects of semions in a supersymmetry  context have been discussed in  \cite{Horvathy:2006pw}.

There is a further special feature of semions in the context of $\mathcal{N}=1$ 3D supersymmetry.  To explain it,  we first recall that 
massive unitary irreps of the 3D Poincar\'e group  are characterized by their  two Poincar\'e invariants \cite{Binegar:1981gv}:  mass and ``relativistic helicity''.  The latter may  be negative; its absolute value is what we shall call spin,  and we abbreviate ``relativistic helicity'' to  ``helicity''.  Parity flips the sign of helicity  so each  state of helicity $h$ is paired, in a parity-preserving theory, with a state of helicity $-h$ with the same mass and quantum numbers. A supermultiplet contains helicities $(h-\tfrac{1}{2},h)$ while its parity-dual contains helicities $(-h, -h +\tfrac{1}{2})$. Generically, these two multiplets are distinct but they are the {\it same} multiplet when $h=\tfrac{1}{4}$.  This suggests that spin-$\tfrac{1}{4}$ particles might arise  naturally in the context of supersymmetric 3D theories preserving parity.  Indeed, the parity self-dual semion supermultiplet of helicities $(-\tfrac{1}{4}, \tfrac{1}{4})$ is known to occur in  various models with two supersymmetry charges, such as ${\cal N}=1$ 3D gauge theories \cite{Witten:1999ds} and  models of supersymmetric quantum mechanics \cite{Pedder:2008je}. The same  supermultiplet, but doubly degenerate,  was recently shown by us to describe  the first excited states of  the 3D ${\cal N}=1$ superstring \cite{Mezincescu:2010yp}. That result led us to a closer inspection of  3D superparticle models. We shall show in a separate paper dedicated to this topic that the generic massive ${\cal N}=1$ supersymmetric parity-preserving superparticle describes, upon quantization, the $(-\tfrac{1}{4}, \tfrac{1}{4})$   semion supermultiplet. 

In this paper we shall be concerned with models that have four supercharges. An example is the 3D abelian-Higgs model with ${\cal N}=2$ supersymmetry.
We shall show that the  Bogomolnyi vortex  \cite{Bogomolny:1975de} of this model is, in the quantum theory,  a  semion of spin $\tfrac{1}{4}$.  It has been previously argued that  vortices in superfluid films are  semions \cite{Chiao:1985zz}; this was on the basis of an analysis of the effective non-relativistic action for a 
system of two vortices, but a subsequent  analysis of the same two-vortex system\cite{Leinaas:1988zz}  concluded that the statistical phase is arbitrary.  Subsequently, field-theoretic models were found with solitons that are naturally semionic \cite{Balachandran:1988ia,Volovik,Salomaa:1987zz}  but these models break parity,  whereas the model considered here is parity-preserving.

Our argument may be summarized as follows. The space of moduli of a single Bogomolnyi vortex solution of the 3D abelian-Higgs model is $\bC$ \cite{Weinberg:1979er}, i.e. position in 2-space viewed as the complex plane.  In the context of the ${\cal N}=2$ supersymmetric extension of  3D abelian-Higgs model,  the Bogomolnyi vortex solution preserves half the supersymmetry  \cite{Edelstein:1993bb}, and therefore breaks only half  the supersymmetry. The resulting fermionic  Nambu-Goldstone modes lead to an extension of the moduli space to the complex superplane  $\bC^{(1,1)}$; we refer to \cite{Tong:2008qd} for a comprehensive review of the physics of vortices in supersymmetric theories.  The low-energy dynamics of  a single vortex is therefore described by an ${\cal N}=2$ massive 3D superparticle action.  This is in the spirit of Manton's moduli space approximation to soliton collisions \cite{Manton:1981mp} but things are simpler for a single soliton: there is no need  to consider only non-relativistic velocities, so the Lorentz invariance of the parent field theory is relevant, as are the supersymmetries.  Taking all (super)symmetries into account leads to a unique effective action governing the dynamics of a single isolated vortex. Of course, the ``dynamics'' is  trivial but that is precisely what allows us to focus on the quantum effects of the fermion  zero  modes. These imply that the quantum states belong to a centrally charged ${\cal N}=2$ parity self-dual semion supermultiplet with helicity content 
\be
(-\tfrac{1}{4},-\tfrac{1}{4}, \tfrac{1}{4},\tfrac{1}{4})\, . 
\ee
As all states have spin $1/4$,  we conclude that the Bogomolnyi vortex has spin $1/4$. 

We shall begin by reviewing  some of the essential features of the Bogomolnyi vortex in the context of the ${\cal N}=2$ supersymmetric  extension of the 
abelian-Higgs model. We then review the light-cone quantization of the massive 3D particle; this serves to introduce some notation that will be useful later. 
Next, we construct  the ${\cal N}=2$ massive superparticle model, which is essentially a 3D variant of the 4D massive superparticle model of 
\cite{Azcarraga:1982dw}. We quantize it in the light-cone gauge to show that it describes  the above  ${\cal N}=2$ semion supermultiplet.
This result assumes that we quantize preserving  parity. We show that the option to quantize breaking parity, which exploits an operator ordering ambiguity, is equivalent to the quantization of a parity-violating extension of the superparticle action to include a ``Lorentz-Wess-Zumino'' term, first 
discussed for the bosonic particle by Schonfeld \cite{Schonfeld:1980kb}.  We conclude with a summary and a discussion of some extensions of our  results.

%%%%%%%%%%%%%%%%%%%%%%%%%%%%%%%
%%%%%%%%%%%%%%%%%%%%%%%%%%%%%%%%

\section{The supersymmetric Bogomolnyi vortex}
\setcounter{equation}{0}

The ${\cal N}=2$ 3D abelian gauge vector multiplet contains the boson fields $(A_\mu, \phi, D)$ where $\phi$ is a real physical scalar and $D$ a real auxiliary scalar, and $A$ is the gauge potential 1-form, with field strength $F=dA$. There is also a complex spinor field $\lambda$.   In the ${\cal N}=2$ supersymmetric 3D abelian-Higgs model this supermultiplet is coupled to  a complex scalar supermultiplet with bosonic fields $(Z, F)$, where $F$ is auxiliary, and  complex spinor field $\chi$.  After elimination of the auxiliary fields, the bosonic truncation of the Lagrangian density is 
\be
\mathcal{L} = -\frac{1}{4}|F|^2  - \frac{1}{2}|\partial\phi|^2
- \left(\partial Z + iq AZ\right) \cdot  \left(\partial Z + iq AZ\right)^* - V(Z,\phi)\, , 
\ee
where $q$ is non-zero constant,  and
\be
V(Z,\phi) = \frac{1}{2} \left( q|Z|^2 - \xi\right)^2 + q^2\phi^2|Z|^2\, , 
\ee
The constant $\xi$ is the Fayet-Iliopoulos parameter. 
If $\xi/q <0$  then the potential cannot vanish and supersymmetry is spontaneously broken. If  
$\xi/q>0$ then there are supersymmetric vacua at 
\be
\phi=0\, , \qquad |Z|= \sqrt{\xi/q}\, . 
\ee
This is the case of interest to us here.  We will not need the remainder of the Lagrangian involving the spinor fields $\lambda$ and $\chi$ but to determine which bosonic solutions preserve some fraction of the supersymmetry of the vacuum solutions we will need to know the supersymmetry variations of these fields, which are 
\begin{eqnarray}\label{fermvars}
\delta_\epsilon \lambda &=& \left[\frac{1}{2} \Gamma^{\mu\nu} F_{\mu\nu} -i \Gamma^\mu \partial_\mu\phi + i \left(q|Z|^2-\xi\right) \right] \epsilon \nonumber \\
\delta_\epsilon \chi &=& \sqrt{2}\left(\partial_\mu Z + i qA_\mu Z\right)  \Gamma^\mu\epsilon \, , 
\end{eqnarray}
where $\Gamma^\mu$ are the 3D Dirac matrices, and $\Gamma^{\mu\nu} = \Gamma^{[\mu} \Gamma^{\nu]}$.

Because the phase of $Z$ is undetermined in a  supersymmetric vacuum, we expect vortex solutions in which this phase changes by some multiple of $2\pi$ as one moves around a circle ``at infinity''. To find them we may set $\phi\equiv 0$. The Hamiltonian density for static configurations then reduces to 
\be
\mathcal{H} = \frac{1}{2} B^2 + \hat \bfn Z \cdot \left(\hat \bfn  Z\right)^* + \frac{1}{2} \left( q|Z|^2 -\xi\right)^2
\ee
where the boldface indicates a spatial 2-vector, and 
\be
B= \bfn \times {\bf A} \, , \qquad \hat\bfn = \bfn +iq{\bf A}\, . 
\ee 
The integral of $\mathcal{H}$ over the 2-space can be written as
\be
%\int \! d^2 x\,  \mathcal{H} = 
\int d^2 x \left\{ \frac{1}{2} \left[ B \pm \left(q|Z|^2 -\xi\right)\right]^2 + \left| \left(\hat\nabla_1 \pm i \hat\nabla_2 \right)Z\right|^2\right\}
\mp \xi \oint_\infty {\bf dx}\cdot {\bf A}
\ee
where the line integral is taken around the circle ``at infinity''; it equals the total magnetic flux of the vortex, and is a topological charge. We therefore have 
the Bogomolnyi bound 
\be 
\int\! d^2 x\,  \mathcal{H} \ge \left| \oint_\infty d{\bf x} \cdot {\bf A} \right| \equiv |m|\, . 
\ee
This inequality is saturated by configurations satisfying the first-order Bogomolnyi equations
\be\label{first-order}
B= \pm \left(q|Z|^2 -\xi\right) \, , \qquad \hat \nabla_1 Z = \pm \,i \hat\nabla_2 Z\, .
\ee
These equations have no known exact solution but a solution exists and can be found numerically.

In a vortex background satisfying (\ref{first-order}) the supersymmetry transformations (\ref{fermvars})  reduce to 
\be
\delta_\epsilon\lambda = \pm iB \left(1\mp i\Gamma_{12}\right)\epsilon \, , \qquad 
\delta_\epsilon \chi = \sqrt{2} \, \Gamma^1 \hat\nabla_1 Z  \left(1\mp i\Gamma_{12}\right)\epsilon\, . 
\ee
Both variations vanish when $i\Gamma_{12}\, \epsilon = \pm \epsilon$, which is equivalent to 
\be
\left[1 \pm i\Gamma^0\right] \epsilon = 0\, . 
\ee
Since $i\Gamma^0$ is hermitian, with eigenvalues $+1$ and $-1$, this condition on $\epsilon$ leaves one of the two complex components of $\epsilon$  unconstrained, which implies that the vortex preserves half of the supersymmetry.  We may go further: this preservation of supersymmetry is possible only 
because of a central  charge in the algebra of the complex supersymmetry charge $Q$. In the presence of the vortex, the non-zero anticommutator of supersymmetry charges in the rest frame must be
\be\label{ftac}
\left\{Q, Q^\dagger \right\} = \left(1 \pm  i\Gamma^0\right) |m| \, . 
\ee

%%%%%%%%%%%%%%%%%%%%%%%%%%
%%%%%%%%%%%%%%%%%%%%%%%%%

\section{Interlude: the 3D particle}
\setcounter{equation}{0}

Let ${\bX}^\mu$  be cartesian coordinates for  3D Minkowski spacetime, with metric $\eta$ of ``mostly plus'' signature. The worldline of a particle in this spacetime, parametrized by a time coordinate $\tau$,  is specified by a map ${\bX}^\mu(\tau)$ from spacetime to the worldline.  We wish to consider an action functional of this map that is invariant under reparametrizations of the worldline. The standard action for a spinless 3D particle of mass $m$ is 
\be\label{particleaction}
S[\bX] = -m \int d\tau \sqrt{\dot {\bX}{}^\mu\dot{\bX}{}^\nu \eta_{\mu\nu}}  \qquad (\mu=0,1,2), 
\ee
The Hamiltonian form of the action is 
\be\label{hamform}
S[\bX,\bP] = \int d\tau \left\{ \dot{\bX}^\mu \bP_\mu - \frac{1}{2}\ell\left(\bP^2  + m^2\right)\right\} \, , \qquad \bP^2 = \eta_{\mu\nu}\bP^\mu\bP^\nu\, . 
\ee
The time reparametrization invariance is equivalent to gauge  invariance under the  infinitesimal ``$\alpha$-symmetry'' transformation  
\be\label{alphasym}
\delta_\alpha {\bX}^\mu = \alpha(\tau){\bP}_\mu\, , \qquad \delta_\alpha {\bP}_\mu =0\, , \qquad 
\delta_\alpha \ell = \dot \alpha\, ,  
\ee
with arbitrary parameter $\alpha(\tau)$. 
This action is also invariant under the rigid action of the 3D Poincar\'e group; the Noether charges are
\be\label{poinc}
\mathcal{P}_\mu = \bP_\mu \, , \qquad \mathcal{J}^\mu =\left[{\bX}\wedge {\bP}\right]^\mu\, , 
\ee
where the wedge product is defined for any two 3-vectors $({\bU},{\bV})$ as
\be\label{wedge}
\left[{\bU}\wedge{\bV} \right]^\mu = \varepsilon^{\mu\nu\rho} {\bU}_\nu{\bV}_\rho\, , \qquad \varepsilon^{012}=1\, . 
\ee

To quantize, we must deal with the gauge invariance. The simplest procedure is to note that the mass-shell constraint is ``first-class'' in Dirac's terminology, and so may be imposed as a physical state condition on the Hilbert space that results from a standard canonical quantization  ignoring  the gauge invariance. 
This leads directly to the Klein-Gordon equation for a spin zero particle of mass $m$.  This procedure is typically much more involved for superparticles, 
and it is generally easier  to first fix the gauge to arrive at an action in canonical form for physical variables only.  We shall illustrate the procedure using the  light-cone gauge. 

We first define  `light-cone'  coordinates and their conjugate momenta  by
\be
x^\pm = \frac{1}{\sqrt{2}}\left({\bX}^1 \pm {\bX}^0\right) \, , \quad x = {\bX}^2  \, ; \qquad 
p_\pm = \frac{1}{\sqrt{2}}\left({\bP}_1 \pm {\bP}_0\right)\, , \quad p= {\bP}_2\, , 
\ee
so that, for example, 
\be
\bP^2 = 2p_- p_+ + p^2\, . 
\ee
We may fix the time reparametrization invariance, equivalently ``$\alpha$-symmetry'' gauge invariance,   by choosing
\be
x^+ =\tau\, . 
\ee
This is the light-cone gauge. In this gauge the Hamiltonian is $H=-p_+$, which  the constraint determines in terms of the remaining canonical pairs. The
resulting Lagrangian is 
\be\label{lcg1}
L= \dot x p + \dot x^- p_- - H\, , \qquad H = \frac{1}{2p_-} \left(p^2+m^2\right)\, . 
\ee
In this gauge, the Poincar\'e charges  (\ref{poinc})  are 
\begin{eqnarray}\label{lcg-poinc}
\mathcal{P} &=& p\, , \qquad \mathcal{P}_- = p_- \, , \qquad \mathcal{P}_+ = -H \, , \nonumber \\
\mathcal{J} &=& x^- p_- + \tau H\, , \qquad \mathcal{J}^+ = \tau p -x p_-\, , \qquad \mathcal{J}^- =  -x^- p - xH\, . 
\end{eqnarray}
The equations of motion imply that these  charges are time-independent; the explicit time-dependence is canceled by the implicit time-dependence due to by the equations of motion.  Upon quantization we have the equal-time commutation relations
\be\label{etcr}
[x^-,p_-] = i\, , \qquad [x,p] =i\, . 
\ee
There are now operator ordering ambiguities in the expressions for $\mathcal{J}$ and $\mathcal{J}^-$. These ambiguities are fixed by the twin requirements of 
hermiticity and   closure of the Lorentz algebra. The quantum Lorentz generators are
\be
\mathcal{J} = \frac{1}{2}\left\{x^-, p_-\right\} + \tau H \, , \qquad \mathcal{J}^+ = \tau p -x p_-\, , \qquad \mathcal{J}^- =  -x^- p - \frac{1}{2}\left\{x,H\right\}\, . 
\ee
It should now be understood that the canonical variables in these expressions are operators, as is $H$.  Using the commutation relations (\ref{etcr}), 
one may verify that 
\be\label{Lcom}
\left[\mathcal{J}, \mathcal{J}^\pm\right]  = \pm i \mathcal{J}^\pm \, , \qquad \left[\mathcal{J}^+,\mathcal{J}^-\right] = i \mathcal{J}\, , 
\ee
which is equivalent to 
\be
\left[\mathcal{J}^\mu , \mathcal{J}^\nu\right] = i \varepsilon^{\mu\nu\rho} \mathcal{J}_\rho\, . 
\ee
One may similarly verify that
\be
\left[\mathcal{J}^\mu , \mathcal{P}^\nu\right] = i \varepsilon^{\mu\nu\rho} \mathcal{P}_\rho\, , \qquad 
\left[\mathcal{P}_\mu , \mathcal{P}_\nu\right] =0 \, . 
\ee
This confirms the Poincar\'e invariance of the quantum theory.

The two Casimirs of the Poincar\'e  algebra are
\be
\mathcal{P}\cdot \mathcal{P} \equiv -m^2  , \qquad  \mathcal{P} \cdot \mathcal{J} =0\, . 
\ee
We thus confirm that the particle has mass $m$ and zero spin zero. This  conclusion may be further verified by an analysis of the time-dependent Schroedinger equation.  Let $\Psi(p,p_-;\tau)$ be the  wave-function in the momentum representation. Taking into account that $\tau= x^+$, the  Schroedinger equation in this representation  is  equivalent to 
\be
\left(p^2+m^2\right) \Psi = 2i p_- \frac{\partial\Psi}{\partial x^+}\, , 
\ee
where $(p,p_-)$ are here the eigenvalues of the corresponding operators.  After a double Fourier transform, this becomes equivalent to 
the Klein-Gordon equation, with mass $m$,  for configuration space wave-function $\Psi(x,x^+; x^-)$. 

%%%%%%%%%%%%%%%%%%%%%%%%%%

\subsection{The LWZ term}
\label{subsec:LWZ}

The particle action (\ref{particleaction})  is invariant under the discrete  parity transformation
\be\label{parity}
{\bX}^2 \to - {\bX}^2\, , \qquad {\bP}_2 \to - {\bP}_2\, . 
\ee
It is possible to add to the action a parity-violating Lorentz-Wess-Zumino (LWZ) term  \cite{Schonfeld:1980kb}. 
This possibility is based on the observation that the phase-space two form\footnote{We suppress the usual wedge symbol for the exterior product in order to avoid possible confusion with the 3D vector product.} 
\be
\Omega = \frac{1}{2}\left(-{\bP}^2\right)^{-\tfrac{3}{2}} \varepsilon^{\mu\nu\rho}\,  \bP_\mu d\bP_\nu d\bP_\rho
\ee
is both manifestly Poincar\'e invariant and closed.  In fact, 
\be
\Omega = d\omega\, , \qquad \omega= - \frac{\varepsilon^{\mu\nu\rho} \bN_\mu {\bP}_\nu d {\bP}_\rho}{\sqrt{-{\bP}^2}\left({\bN}\cdot {\bP} + \sqrt{-{\bP}^2}\right)} \, , \qquad {\bN}^2=-1\, ,
\ee
where $\bN$ is a constant normalized timelike 3-vector. The pullback of $\omega$ to the worldline defines a worldline 1-form $d\tau\, \omega_\tau$ and we may add to the  action  the term 
\be
L_{LWZ} = \beta\! \int d\tau\,  \omega_\tau\, . 
\ee
Although this is not manifestly Lorentz invariant, its Lorentz variation is a boundary term.  This is sufficient for Noether's theorem but the Lorentz Noether charges are modified to 
\be\label{Lbeta1}
\mathcal{J}^\mu =\left[{\bX}\wedge {\bP}\right]^\mu - \frac{\beta}{m} {\mathcal P}^\mu\, . 
\ee
It follows that
\be\label{shifthel}
\mathcal{P} \cdot \mathcal{J} = m \beta\, , 
\ee
so the effect of the LWZ term is to shift the helicity by $\beta$; this means that for $2\beta \notin \bZ$ the particle is an ``anyon''.  As far as we are aware, this was the first  `physical'  realization  of fractional angular  momentum (in contrast to fractional statistics \cite{Leinaas:1977fm}).

In the quantum theory, the new $\beta$-dependent Lorentz charges (\ref{Lbeta1}) must  be equivalent to the standard Lorentz commutation relations, for any 
$\beta$; this is possible because the canonical commutation relations are also $\beta$-dependent  since they are altered by the addition of the LWZ term.  How this all works  is quite transparent in the light-cone gauge. The addition of the LWZ term leads to the  modified light-cone-gauge Lagrangian
\be
L_\beta = \dot x p + \dot x^- p_- + \beta \frac{p \dot p_-}{m p_-} - H\, ,
\ee
where $H$ is as before.  It is  now possible to redefine variables so as to remove the LWZ term from the action!  Let us define
\be
y^- = x^- - \frac{\beta p}{m p_-} \, , \qquad  y = x + \frac{\beta}{m}\, . 
\ee
The shift of $x$ has no effect on the Lagrangian but it affects the Lorentz charges because these depend on the origin of coordinates, even though the
Poincar\'e invariants obviously do not. The shift of $x^-$ affects both the Lagrangian and the Lorentz charges. The combined effect is that the Lagrangian 
becomes
\be\label{lcg2}
{\cal L} = \dot y p + \dot y^- p_-  -H \, , \qquad H = \frac{1}{2p_-} \left(p^2+m^2\right)\, . 
\ee
This differs from  (\ref{lcg1}) only in  notation. However, the Lorentz charges are now
\be
\mathcal{J} = y^- p_- + \tau H\, , \qquad \mathcal{J}^+ = \tau p -y p_-\, , \qquad \mathcal{J}^- =  -y^- p - yH + \beta \frac{m}{p_-}\, . 
\ee
From this result we recover the result (\ref{shifthel}) for the shift in helicity.  Despite the  addition of $\beta m/p_-$ to $\mathcal{J}^-$, 
the Lorentz  charges are still hermitian and satisfy the commutation relations (\ref{Lcom}).  Thus, the effect of the LWZ term in the light-cone gauge is to 
exploit an ambiguity in the definition of $\mathcal{J}^-$ that shifts  the helicity by a constant.

%%%%%%%%%%%%%%%%%%%%%%%%%
%%%%%%%%%%%%%%%%%%%%%%%%%

\section{The massive ${\cal N}=2$ superparticle}
\setcounter{equation}{0}

A superparticle action is an extension of the particle action just considered in which the spacetime, in our case 3D Minkowski space, is 
extended to a superspace with additional anticommuting coordinates. For the case of interest here, the additional anticommuting coordinates
transform as a single complex 3D spinor $\Theta$, but this is equivalent to a pair of Majorana spinors. The equivalence is especially 
simple if we choose a real  representation of the Dirac matrices such as 
\be
\Gamma_0= -i\sigma_2\, , \qquad \Gamma_1 = \sigma_1\, , \qquad \Gamma_2 = \sigma_3\, . 
\ee
In this case, the real and imaginary parts of the complex spinor $\Theta$ are Majorana spinors:
\be
\Theta=  \Theta_1 + i \Theta_2\, , \qquad \bar\Theta_a = \Theta_a^T \Gamma^0 \qquad (a=1,2). 
\ee
The superspace with coordinates $({\bX}, \Theta_a)$ is the  group manifold for the ${\cal N}=2$ supertranslation group. This includes the supersymmetry transformations, which have the following infinitesimal action on the superspace coordinates:
\be
\delta_\epsilon {\bX}^\mu = i \bar\Theta_a \Gamma^\mu \epsilon_a \, ,  \qquad \delta_\epsilon \Theta_a = \epsilon_a
\ee
where $\epsilon_a$ is a pair of  constant real anticommuting spinor parameters . 

The following superspace differential forms are invariant under supersymmetry transformations 
\be\label{diff1}
\Pi^\mu \equiv d{\bX}^\mu + i \bar\Theta_a \Gamma^\mu d\Theta_a\, , \qquad 
d\Theta_a\, . 
\ee
We may construct a manifestly super-Poincar\'e invariant action from the previously discussed particle action by the replacement
\be
\dot {\bX} \to \Pi_\tau \equiv \dot{\bX} + i \bar\Theta_a \Gamma^\mu \dot \Theta_a\, . 
\ee
However, there will be no central charge in the anticommutator of the supersymmetry Noether charges of this model, precisely because
the super-Poincar\'e symmetry is manifest. To allow for the central charge we must add a Wess-Zumino (WZ) term to the action. To do this we 
observe that the  real 2-form\footnote{We adopt the common convention that the product of a pair of  real anticommuting  variables is imaginary, so the factor of $i$ is needed for reality.} 
\be
i\epsilon^{ab} d \bar\Theta_a d \Theta_b = d\left[i \varepsilon^{ab}  \bar\Theta_a d \Theta_b\right]
\ee
is both super-Poincar\'e invariant and closed. In fact,  it is exact in de Rham co-homology but not in the Lie algebra cohomology of relevance here
because the 1-form $\varepsilon^{ab}  \bar\Theta_a d \Theta_b$ is not supertranslation invariant; its variation under supersymmetry is a total derivative. 
However, this is sufficient for the purposes of constructing a super-Poincar\'e invariant action.  

These considerations lead to the ${\cal N}=2$ superparticle 
action
\be\label{sparticle2}
S[\bX,\bP,\Theta_a; \ell] =  \int\! d\tau \left\{ \dot {\bX}^\mu {\bP}_\mu + i \bar\Theta_a \Pslash \, \dot\Theta_a  - i   m \varepsilon^{ab} \bar\Theta_a \dot\Theta_b
- \frac{1}{2}\ell \left(\bP^2 + m^2\right)\right\}\, . 
\ee
where $\Pslash = {\bP}_\mu \Gamma^\mu$. 
This action is invariant under time reparametrizations, equivalently $\alpha$-symmetry gauge transformations, which are exactly as for the bosonic particle. 
In addition, we have chosen the coefficient of the WZ term to ensure invariance under the following (``$\kappa$-symmetry'') gauge transformations:
\be
\delta_\kappa X^\mu = i\delta_\kappa\bar\Theta_a \Gamma^\mu \Theta_a \, , \qquad 
\delta\Theta_a = \left(\Pslash\,  \delta^{ab}+ m \varepsilon^{ab}\right) \kappa_b
\, , \qquad \delta_\kappa\ell= -4i \bar\kappa_a \dot \Theta_a\, . 
\ee
As half of the eigenvalues of the matrix $(\Pslash\, \delta + m \varepsilon)$ vanish on the mass shell, only half of the components of $\Theta_a$ may be ``gauged away. This means that there is only one complex gauge-invariant combination of the components of  $\Theta_a$. Taking into account the time-reparametrization invariance,  we conclude that the superparticle action (\ref{sparticle2}) describes dynamics on $\bC^{(1,1)}$, as required.

The Lorentz Noether charges for the action (\ref{sparticle2}) are
\be\label{Lorentz2}
\mathcal{J}^\mu = \left[\bX \wedge \bP\right]^\mu + \frac{i}{2} \bar\Theta_a \Theta_a\,  
\bP^\mu - \frac{i m}{2} \varepsilon^{ab} \bar\Theta_a \Gamma^\mu \Theta_b\, , 
\ee
The supersymmetry Noether charges are 
\be\label{N=2susy}
{\cal Q}_a = \sqrt{2}\, \left[ \Pslash\, \delta^{ab} -  m\, \varepsilon^{ab} \right]\Theta_b \, , 
\ee
The $\alpha$-symmetry and $\kappa$-symmetry variations of these  charges vanish on the mass-shell.  Finally, to complete our discussion of the 
symmetries,  we observe that the action  (\ref{sparticle2}) is invariant under the discrete symmetry
\be\label{parity}
{\bX}^2 \to - {\bX}^2\, , \qquad {\bP}_2 \to - {\bP}_2\, , \qquad \Theta_1 \to \Gamma_2 \Theta_1\, , \qquad 
\Theta_2 \to -\Gamma_2 \Theta_2\, . 
\ee
This shows that the  parity invariance of the bosonic particle extends to a symmetry of the ${\cal N}=2$ massive superparticle, {\it even in the presence of the WZ term}. 

%%%%%%%%%%%%%%%%%%%%%%%%%%%%%

\subsection{Light-cone gauge Quantization}

We shall quantize in the light-cone gauge. First, we  fix the $\kappa$-symmetry by imposing the condition 
\be\label{vartheta}
\Gamma^+ \Theta_a =0 \, . 
\ee
This implies that 
\be 
\Theta_1 + i\Theta_2 = \frac{1}{\sqrt{\sqrt{2}\, p_-}} \left(
\begin{array}{c} \theta \\ 0 \end{array} \right)\, , 
\ee
where $\theta$ is a single complex anticommuting variable. The light-cone gauge fixing of the $\alpha$-gauge transformations now proceeds as for the bosonic particle: we set $x^+=\tau$ and  solve the constraint for $p_+$.
The resulting Lagrangian is 
\be\label{lightconesuperlag}
{\cal L} = \dot x p + \dot x^- p_- + i \theta^* \dot\theta  -H \, , \qquad H= \frac{p^2 +m^2}{2p_-}\, . 
\ee

The Poincar\'e generators in the light-cone gauge are 
\begin{eqnarray}\label{lcg-poinc}
\mathcal{P} &=& p\, , \qquad \mathcal{P}_- = p_- \, , \qquad \mathcal{P}_+ = -H \, , \nonumber \\
\mathcal{J} &=& x^- p_- + \tau H\, , \qquad \mathcal{J}^+ = \tau p -x p_-\, , \nonumber \\
 \mathcal{J}^- &=&  -x^- p - xH  + \frac{m}{2 p_-} \theta \theta^*\, , 
\end{eqnarray}
and hence
\be
m^{-1} {\cal P} \cdot {\cal J} =  \frac{1}{2} \theta \theta^*\, . 
\ee
The {\it complex} supersymetry charge 
\be
Q= \mathcal{Q}_1 +i \mathcal{Q}_2 
\ee
becomes, in the light-cone gauge, 
\be
Q= \frac{1}{\sqrt{\sqrt{2}\ p_-}} \left( \begin{array}{c} \left(p+ im\right)\theta \\ \sqrt{2}\, p_- \theta \end{array} \right)\, . 
\ee
Finally, we observe that the parity transformations of (\ref{parity}) imply the following discrete transformations of the light-cone variables:
\be\label{lcgparity}
x \to -x\, , \qquad p\to -p\, , \qquad \theta \to \theta^*\, .
\ee
The Lagrangian (\ref{lcg-poinc}) changes  by a total derivative under this transformation, so the gauge-fixing has not destroyed the parity invariance. 

Upon quantization, which involves $\theta^*\to \theta^\dagger$,  we have the equal-time (anti)commutation relations
\be\label{varthetaantic}
[x^-,p_-] = i\, , \qquad [x,p] =i\, , \qquad  \left\{\theta^\dagger, \theta \right\} =  1\, . 
\ee
The  anticommutation relation implies a two-state system. It also implies that
\be
\left\{ Q, Q^\dagger \right\} = \left(\Pslash + im \right)\Gamma^0\, , 
\ee
where we may interpret the matrix operator  on the right hand side as an ordinary matrix in the momentum representation. 
Going to the rest-frame we then find that
\be
\left\{ Q, Q^\dagger \right\} =\left(1\pm i \Gamma^0\right)|m|\, , 
\ee
where the sign is the sign of $m$. This is precisely the anticommutator of (\ref{ftac}). 

The quantum Lorentz  charges are
\begin{eqnarray}\label{Lorentzbeta}
\mathcal{J} &=& \frac{1}{2}\left\{x^-, p_-\right\} + \tau H \, , \qquad \mathcal{J}^+ = \tau p -x p_-\, , \nonumber \\
 \mathcal{J}^- &=&  -x^- p - \frac{1}{2}\left\{x,H\right\} + \frac{m}{4p_-}\left[\theta,\theta^\dagger\right] + \frac{m\beta}{p_-} \, ,
\end{eqnarray}
where $\beta$ is an arbitrary real constant arising from the ambiguity in ordering of the $\theta$ and $\theta^\dagger$ operators\footnote{This ordering ambiguity does not occur if the classical Lagrangian is first expressed in terms of the real and imaginary parts of $\theta$. Nevertheless, there is still an ambiguity in the definition of the quantum charges.}.  We encountered this ambiguity in subsection \ref{subsec:LWZ} where we saw that it corresponds to the addition to the classical action of a LWZ term. The addition of this term to the superparticle action is manifestly compatible with all gauge invariances, and with super-Poincar\'e invariance, but it breaks parity.  This is obvious from the quantum version of the parity transformations (\ref{lcgparity}), which interchange $\theta$ with $\theta^\dagger$, but it can also be seen from the fact that   it is only for $\beta=0$ that the parity transformations (\ref{lcgparity})  induce an automorphism of the Lorentz algebra:
\be
\mathcal{J}\to \mathcal{J} \, , \qquad \mathcal{J}^\pm \to - \mathcal{J}^\pm \, . 
\ee
As our starting action had no LWZ term, and hence preserved parity, we now set $\beta=0$. With this choice, we have
\be
m^{-1} {\cal P} \cdot {\cal J}  = \frac{1}{4} - \frac{1}{2} F  \, , \qquad F = \theta^\dagger\theta\, . 
\ee
The fermion number operator $F$ has eigenvalues $0$ (empty) or $1$ (filled). The ``empty'' state therefore has helicity $\tfrac{1}{4}$ while the ``filled'' state 
has helicity $-\tfrac{1}{4}$. Moreover,
\be
\left\{ \left(-1\right)^F, Q\right\} =0\, , \qquad \left(-1\right)^F \equiv 1-2F\, , 
\ee
which implies that the ``empty'' and ``filled''  states are part of a supermultiplet; this is possible because the helicity difference is $1/2$. 
Actually, each of the ``empty'' and ``filled'' states is doubly degenerate. This follows from the fact that the following three  hermitian operators are mutually anticommuting:
\be 
\left(\theta+\theta^\dagger\right) \, , \qquad i \left(\theta- \theta^\dagger\right)\, , \qquad \left(-1\right)^F \equiv 1-2\theta^\dagger \theta\, . 
\ee
There is no real hermitian $2\times 2$ realization of three such operators, so  the two-dimensional spin space is necessarily complex. In other words, there are two ``empty''  states of helicity $\tfrac{1}{4}$ and two  ``filled'' states of helicity $-\tfrac{1}{4}$. This is also expected from the fact  that the states must carry a charge.  We thus conclude that the helicity content of the supermultiplet described by the superparticle is 
\be
\left(-\tfrac{1}{4},-\tfrac{1}{4}, \tfrac{1}{4},\tfrac{1}{4}\right)\, . 
\ee
This  parity self-dual `short' representation of ${\cal N}=2$ supersymmetry with a central charge is one that has not previously been considered, to the best of our knowledge.  Each of the two `active' supersymmetry charges takes one of the $-\tfrac{1}{4}$ helicity states to one of the $\tfrac{1}{4}$ helicity states. Usually one expects spacetime supersymmetry to connect  different spins but here is an exception!

%%%%%%%%%%%%%%%%%%%%%%
%%%%%%%%%%%%%%%%%%%%%

\section{Discussion} 

We have shown that the $\kappa$-symmetric 3D massive ${\cal N}=2$ superparticle describes, upon quantization and preserving parity invariance, a supermultiplet of states, all of which have spin $1/4$.  As far as we are aware, this remarkable supermultiplet of ${\cal N}=2$ 3D supersymmetry was previously unknown\footnote{An $\mathcal{N}=1$ semion supermultiplet with a ``hidden'' $\mathcal{N}=2$ supersymmetry was described in \cite{Gorbunov:1997ie} but it is not clear to us what the action of the central charge is in this construction.}. It would of interest to find a Lorentz covariant description of it, perhaps along the lines of \cite{Sorokin:1992sy}. 

Equally remarkable is the fact that this superparticle action  can be interpreted as an effective action for a  Bogomolnyi vortex of the ${\cal N}=2$ supersymmetric 3D abelian-Higgs model; this interpretation shows that the quantum Bogomolnyi vortex in this context carries spin $\tfrac{1}{4}$.   
Our result may well  apply  to other supersymmetric soliton solutions of parity-preserving 3D field theories with ${\cal N}=2$ supersymmetry, but in general one expects additional ``internal'' degrees of freedom to be relevant. In this case it might be possible to view the spin-$\tfrac{1}{4}$ soliton as a bound state of some more fundamental ingredients. 

Our results may be extended to ${\cal N}>2$. For even ${\cal N}$ there is a kappa-symmetric massive superparticle action. For ${\cal N}=4$ it describes a supermultiplet of helicities 
\be
\left(-\tfrac{1}{2},-\tfrac{1}{2}, 0,0,0,0, \tfrac{1}{2},\tfrac{1}{2}\right)\, . 
\ee
This is just the 3D version of the 4D hypermultiplet. This case is of course applicable to the Bogomolnyi vortex of the maximally-supersymmetric ${\cal N}=4$ abelian-Higgs model, but in this context the result is not a surprise. 

For ${\cal N}=6$ we again find a semion supermultiplet with helicities
\be
\left( -\tfrac{3}{4},-\tfrac{3}{4}, -\tfrac{1}{4},-\tfrac{1}{4},-\tfrac{1}{4}, -\tfrac{1}{4},-\tfrac{1}{4}, -\tfrac{1}{4},\tfrac{1}{4}, \tfrac{1}{4},\tfrac{1}{4}, \tfrac{1}{4}, \tfrac{1}{4},
\tfrac{1}{4},\tfrac{3}{4},\tfrac{3}{4}\right)\, . 
\ee
However, there is no obvious application to 3D field theory. At ${\cal N}=16$ we get a parity self-dual supermultiplet with spin 2. This should not be confused with the recently discussed  ${\cal N}=8$ parity self-dual spin 2 supermultiplet \cite{Bergshoeff:2010ui} because the supermultiplets under discussion here have a central charge. 

Finally, we have clarified some aspects of the parity-violating Lorentz-Wess-Zumino term introduced for the 3D particle in \cite{Schonfeld:1980kb}. This
is perhaps the simplest mechanical model for an anyon. In particular, we have pointed out that this term can be included in superparticle models, consistent with all gauge invariances and all symmetries except parity. In this context it leads to a shift of all helicities in a supermultiplet. Applied to the massive ${\cal N}=2$ superparticle, it would be possible to choose the coefficient of the LWZ term to arrange for the helicities to 
be $(0,0, \tfrac{1}{2},\tfrac{1}{2})$; i.e. non-anyonic. The absence of helicity $-\frac{1}{2}$ states in this supermultiplet shows that parity is broken.
What is not clear is how the LWZ term could arise in an effective action for vortices  in the parity-violating extension of  the  ${\cal N}=2$ supersymmetric abelian-Higgs model to include a Chern-Simons term. The results of \cite{Collie:2008mx} suggest that Chern-Simons terms have a rather different effect, but it is also not clear to us at present how this effect is compatible with supersymmetry.  It seems that there remains much to be understood about solitions in 
parity-violating  supersymmetric 3D field theories.

\bigskip
\noindent
\section*{Acknowledgements}
We thank  the Benasque center for Science, where this paper  was written, for a stimulating environment. We are also grateful to David Tong for helpful correspondence.   LM acknowledges partial support from National Science Foundation Award 0855386.  PKT thanks the  EPSRC for financial support.

\end{document}